\documentclass[11pt]{article}
\usepackage{cite}

\usepackage{hyperref}
\usepackage[onehalfspacing]{setspace}
\usepackage{graphicx}% Include figure files
\usepackage[left=2.5cm,right=2.5cm,bottom=3cm,includeheadfoot,a4paper]{geometry}

\newcommand{\bra}[1]{\langle {#1} |}
\newcommand{\ket}[1]{| {#1} \rangle}
\newcommand{\braket}[2]{\langle {#1} | {#2} \rangle}
\newcommand{\FN}{F_{N}}

\begin{document}
%% \doi{10.1080/1478643YYxxxxxxxx}
%% \issn{1478-6443}
%% \issnp{1478-6435}
%% \jvol{00} \jnum{00} \jyear{2010} \jmonth{00}

\markboth{R.~Lifshitz and S.~Even-Dar Mandel}{Log-periodic oscillations in
  Fibonacci quasicrystals} 

\title{Observation of Log-Periodic Oscillations in the Quantum
  Dynamics of Electrons on the One-Dimensional Fibonacci Quasicrystal}

\author{Ron Lifshitz and Shahar Even-Dar Mandel\\
  Raymond and Beverly Sackler School of Physics and Astronomy\\
  Tel Aviv University, Tel Aviv 69978, Israel} 
%\date{today}
\date{June 3, 2010}
\maketitle

\begin{abstract}
  
  We revisit the question of quantum dynamics of electrons on the
  off-diagonal Fibonacci tight-binding model. We find that typical
  dynamical quantities, such as the probability of an electron to
  remain in its original position as a function of time, display
  log-periodic oscillations on top of the leading-order power-law
  decay. These periodic oscillations with the logarithm of time are
  similar to the oscillations that are known to exist with the
  logarithm of temperature in the specific heat of Fibonacci
  electrons, yet they offer new possibilities for the experimental
  observation of this unique phenomenon.

%% \begin{keywords}
%%   Quasicrystals, Fibonacci quasicrystals, tight-binding model,
%%   electronic transport, quantum dynamics, log-periodic oscillations
%% \end{keywords}
\end{abstract}

\section{Fibonacci Electrons}

The Fibonacci sequence of \emph{Long} ($L$) and \emph{Short} ($S$)
intervals on the 1-dimensional line---generated by the simple
substitution rules $L \to LS$ and $S \to L$---is a favorite textbook
model for demonstrating the peculiar nature of electrons in
quasicrystals\cite{review1,review2,review4}. The wave functions of
Fibonacci electrons are neither extended nor exponentially-localized,
but rather decay algebraically; the spectrum of energies is neither
absolutely-continuous nor discrete, but rather singular-continuous,
like a Cantor set; and the quantum dynamics is anomalous. In recent
years we have studied how these three electronic properties change as
the dimension of the Fibonacci quasicrystal increases to two and
three~\cite{ilan,shahar06,shahar08,bloch,fibdynamics}, by constructing
square and cubic versions of the Fibonacci
quasicrystal~\cite{squarefib}.

The 1-dimensional \emph{off-diagonal} Fibonacci tight-binding model is
constructed by associating a unit hopping amplitude between sites
connected by a \emph{Long} interval, and a hopping amplitude $T>1$
between sites connected by a \emph{Short} interval, while assuming
equal on-site energies that are taken to be zero. The resulting
tight-binding Schr\"odinger equation, on an $\FN$-site model, is given
by
\begin{equation}
  \label{eq:schrodinger}
  T_{j+1} \psi(j+1) + T_j \psi(j-1) = E \psi(j),\quad
  j=1\ldots \FN,
\end{equation}
where $\psi(j)$ is the value of an electronic eigenfunction on site
$j$, $E$ is the corresponding eigenvalue, $\FN$ is the $N^{th}$
Fibonacci number, and the hopping amplitudes $T_j$ are equal to 1 or
$T$ according to the Fibonacci sequence $\{T_j\} =
\{1,T,1,1,T,1,T,1,1,T,1,1,T,1,T,1,1,T, 1,T,\ldots\}$, as described
above. The \emph{diagonal} Fibonacci tight-binding model is
constructed by taking all the hopping amplitudes to be equal to 1, and
taking the on-site energies to have two values, $V_L$ and $V_S$,
arranged according to the Fibonacci sequence. These models have been
studied extensively ever since the initial interest in the behavior of
electrons in quasiperiodic
potentials~\cite{1dmodel1,1dmodel2,1dmodel3,1dmodel4}, and continue to
offer mathematical challenges to this day~\cite{damanik2008}.

Various 2-dimensional extensions of the Fibonacci model were
introduced soon
thereafter~\cite{2dmodel1,2dmodel2,2dmodel3,2dmodel4,2dmodel5}, and
strongly promoted recently~\cite{squarefib} as models for
quasicrystals without `forbidden'
symmetries~\cite{definition1,definition2}. In our studies, we have
shown that whereas Fibonacci electrons in one dimension always behave
as described above for any $T>1$, in two dimensions, and even more so
in three, there is crossover---as the strength $T$ of the
quasiperiodicity approaches 1---to a regime in which Fibonacci
electrons behave more and more like electrons do in periodic crystals,
particularly in the sense that their energy spectra develop continuous
intervals~\cite{ilan,shahar06,shahar08}. These results were recently
explained in a rigorous manner by Damanik and
Gorodetzki~\cite{damanik2009}. More surprisingly, our studies of
Fibonacci electrons have led us to new results also in the simple
1-dimensional case. We have examined dynamical properties, such as the
probability of an electron to remain in its original position as a
function of time.  The power-law decay of this quantity is commonly
used for analyzing the dynamics. Surprisingly, we have observed
\emph{log-periodic oscillations} on top of the power-law decay,
implying the existence of an imaginary correction to the exponent. We
wish to describe this new observation here.

\section{Quantum Dynamics of Electronic Wave-Packets}
\label{sec:dynamics}

We consider the dynamics of electronic wave-packets, or states
$\ket{n}$, that are initially localized at a single lattice site $n$
of the 1-dimensional Fibonacci quasicrystal, denoting their amplitude
on site $m$ at time $t$ by $\phi_n(m,t)$. Thus, at time $t=0$ the
wave-packet is given by $\phi_n(m,0)=\braket{m}{n}=\delta_{mn}$, where
$\delta_{mn}$ is the Kronecker delta. At any later time $t>0$ the
wave-packet is given by $e^{iHt}\ket{n}$, which is simply the $n^{th}$ column of the matrix
representation of the time evolution operator $e^{iHt}$, where $H$ is
the off-diagonal matrix representation of the Hamiltonian in
Eq.~(\ref{eq:schrodinger}), and we take $\hbar=1$. Thus,
$\phi_n(m,t)= \bra{m} e^{iHt}\ket{n} = (e^{iHt})_{mn}$.

We characterize the dynamics of wave packets by monitoring two typical
quantities: (a) The \emph{Survival Probability} of the $n^{th}$ wave-packet,
defined as the probability of finding the electron at its initial
position at time $t$, is given by
\begin{equation}\label{eq:survdef}
S_n(t)=|\phi_n(n,t)|^2 = \left|(e^{iHt})_{nn}\right|^2;
\end{equation}
and (b) The \emph{Inverse Participation Ratio} of the $n^{th}$
wave-packet, which measures the spatial extent of the wave-packet, is
given by
\begin{equation}\label{eq:iprdef2}
I_n(t) = \frac{\sum_m \left|\phi_n(m,t)\right|^4}{\left(\sum_m
    \left|\phi_n(m,t)\right|^2\right)^2} =
\sum_m \left|(e^{iHt})_{mn}\right|^4,
\end{equation}
where the last equality holds because the wave-packets are
normalized. Both of these quantities are often used to examine the
dynamics of wave-packets. In particular, the manner in which they
decay as the wave-packets spread with time is associated with
different regimes of the quantum dynamics. The actual calculation is
performed numerically by applying the discrete time-evolution operator
$e^{iH\Delta t}$ successively to the initial conditions. The time-step
$\Delta t$ is kept sufficiently small to satisfy the Nyquist criterion
by noting that the largest eigenvalue of $H$ is bounded by $1+T$, thus
requiring $\Delta t$ to be smaller than $1/(2+2T)$.

%For an infinitely-sized system the time can be taken to infinity
%without getting the results affected by boundary conditions. 

If a function $F(t)$ decays asymptotically with some power law
$F(t)\sim t^{-\beta}$, then the exponent $\beta$ can be found by
\begin{equation}
\beta=\lim_{t\rightarrow\infty}-\frac{\ln F(t)}{\ln t},
\end{equation}
but in general the exponent $\beta$ is not guaranteed to exist. The
bounds $\beta_{\pm}$ on $\beta$, which always exist, are given by
\begin{equation}\label{eq:defbetas}
\beta_{+}=-\limsup_{t\rightarrow\infty}\frac{\ln F(t)}{\ln
  t};\quad \beta_{-}=-\liminf_{t\rightarrow\infty}\frac{\ln
  F(t)}{\ln t}; 
\end{equation}
and if $\beta_{+}=\beta_{-}$ then $\beta$ exists. Furthermore, if an
exponent $0<\beta<1$ characterizes the power-law decay of a function
$F(t)$, then the same exponent also characterizes the decay of the
time-averaged function
\begin{equation}\label{eq:timeav_def}
\langle F \rangle_t=\frac{1}{t}\int_0^t F(t') {\rm d}t'.
\end{equation}
We use the definition of Eq.~(\ref{eq:defbetas}) to study the
long-time asymptotic values of the exponents $\beta_S$ related to the
survival probability of a wave-packet and $\beta_I$ related to the
inverse participation ratio of the wave-packet. In what follows we
are mostly concerned with the early-time behavior of these exponents.

Damanik \emph{et al.}~\cite{damanik2008,damanik2010} recently used the
second moment of the position operator for the diagonal tight-binding
Hamiltonian, to show that far from the periodic limit, or very close
to it, the dynamics of wave-packets is independent of the initial
site. However, since we are interested in intermediate values of $T$
as well---equivalent in the diagonal model to intermediate values of
the difference $|V_L - V_S|$ between the different on-site
energies---we expect to find different dynamical behavior depending on
the choice of the initial site. Thus, we typically examine the
maximal, the minimal, and the site-averaged survival probabilities and
inverse participation ratios, all of which display similar qualitative
behavior. In Ref.~\cite{fibdynamics} we study the different exponents
in detail, in one, two, and three dimensions. Here we concentrate on
the $1d$ results for the maximal survival probability exponent
$\beta_S^{max}$.

\section{Asymptotic Behavior of the Maximal Survival Probability}
\label{sec:extremaldynamics}

\begin{figure}
\begin{center}
\scalebox{0.7}{\includegraphics{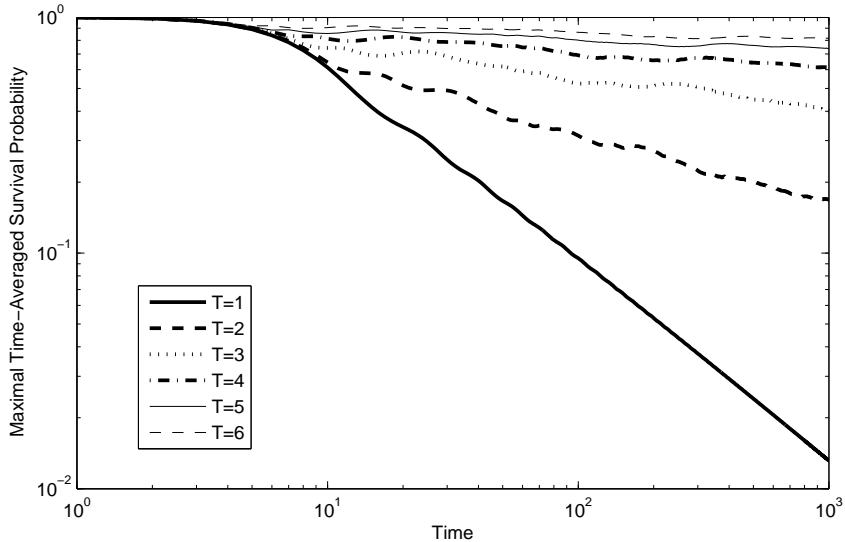}}
\caption{\label{fig:MaxSTimeAv} The time-averaged maximal survival
  probability for a 233-site $1d$ approximant, with periodic boundary
  conditions, calculated for different values of $T$. The asymptotic
  behavior of the slopes allows to extract the diffusion exponent
  $\beta_S^{max}$ as a function of $T$, which is displayed in
  Fig.~\ref{fig:pMaxSTimeAv}.}
\end{center}
\end{figure}

\begin{figure}
\begin{center}
\scalebox{0.7}{\includegraphics{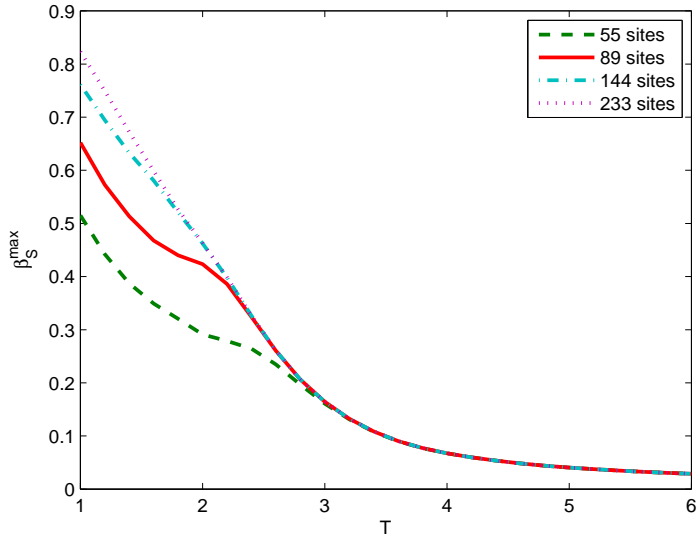}}
\caption{\label{fig:pMaxSTimeAv} The exponent $\beta_S^{max}$ as a
  function of $T$ for increasing orders of approximants. Convergence
  is improved for decreasing values of $T$ as the order of the
  approximant is increased.}
\end{center}
\end{figure}

We look at the wave-packet whose time-averaged survival probability is
maximal. Fig.~\ref{fig:MaxSTimeAv} shows the time-averaged maximal
survival probability $\langle S_{max} \rangle_t$ on a log-log scale
for a 233-site $1d$ approximant, with periodic boundary conditions,
for different values of $T$.  The exponents extracted from the slopes
of the curves in Fig.~\ref{fig:MaxSTimeAv} for four orders of
approximants are shown in Fig.~\ref{fig:pMaxSTimeAv}, as functions of
$T$. 

Convergence of the exponent $\beta_S^{max}$ is evident for values of
$T>2$ even for relatively small approximants. Convergence is not
obtained for smaller values of $T$ that approach the periodic limit
$T\to 1$, where the dynamics is expected to become ballistic. In this
limit the wave-packets quickly spread out and the finite size of the
approximant has a stronger influence on the dynamics.  Despite this
limitation, detailed studies of the slopes, for increasing orders of
approximants and for varying time-ranges clearly show that the
extracted exponents converge, as the order of the approximant
increases, and longer averaging times are possible. We note that
Similar curves are obtained for the minimal and the site-averaged
survival probabilities, as well as for the minimal, maximal, and
site-averaged inverse participation ratios~\cite{fibdynamics}.

\section{Log-Periodic Oscillations and the Fourier Transform of the
  Density of States}
\label{sec:logperiodic}

\begin{figure}
\begin{center}
\scalebox{0.75}{\includegraphics{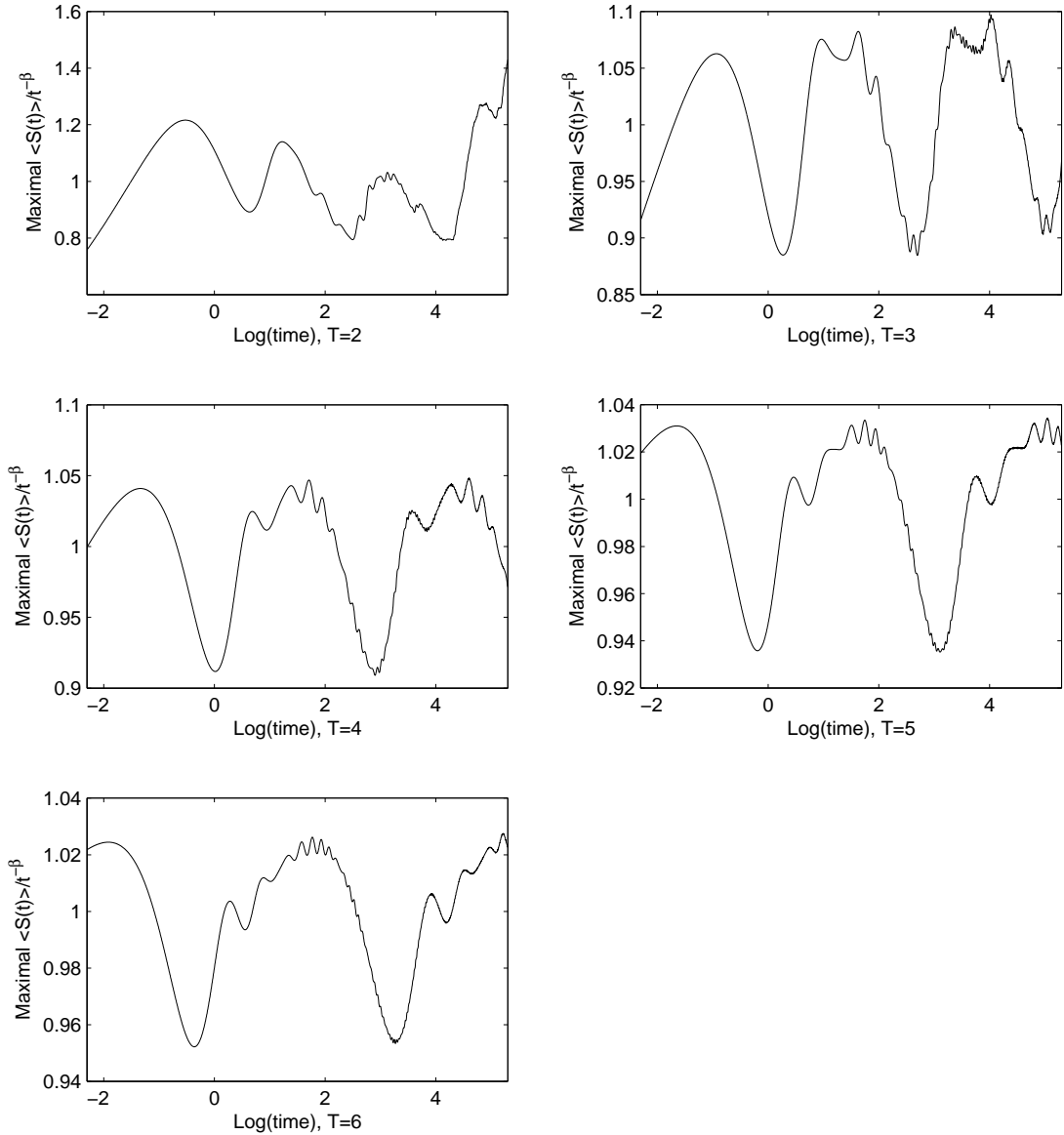}}
\caption{\label{fig:logperiodic} Log-periodic oscillations in the
  maximal survival probability, after dividing out the leading
  power-law decay, for an 89-site approximant with 5 different values
  of $T$. In addition to the basic periodic oscillation one observes
  the emergence of higher-frequency oscillations as time advances.}
\end{center}
\end{figure}

A closer inspection of the temporal decay of the survival probabilities
in Fig.~\ref{fig:MaxSTimeAv} reveals a small oscillating behavior on
top of the overall leading power-law. In order to better expose this
behavior, we divide out the leading asymptotic power-law $t^{-\beta}$
that was found earlier, and obtain the curves shown in
Fig.~\ref{fig:logperiodic}. Similar results are observed in the study
of the inverse participation ratio. The curves clearly exhibit
log-periodic oscillations---oscillation that are periodic in
$\log(t)$---around the mean value of 1, especially for the larger
values of $T$. These oscillations contain a basic frequency $\omega$
that seems to decrease with increasing $T$, as well as a sequence of
higher-frequency oscillations with decreasing amplitudes that seem to
develop with time. The fundamental oscillations can be described
empirically by a temporal decay with a complex exponent of the form
\begin{equation}
f(t)\propto t^{-\beta} + \alpha t^{-\beta\pm i\omega} + ({\rm
  corrections} \ll\alpha), 
\end{equation}
which after dividing out the leading power-law term become
\begin{equation}
\frac{f(t)}{t^{-\beta}} \propto 1+\alpha t^{\pm i\omega}=1+\alpha
e^{\pm i\omega\log(t)}, 
\end{equation}
yielding oscillations of amplitude $\alpha$ around 1 that are periodic
in $\log(t)$ with frequency $\omega$.

Log-periodic oscillations are not as uncommon in physics as one may
think. They appear in critical
phenomena~\cite{nauenberg75,niemeijer76,jullien81}, where
renormalization-group calculations yield power-law behavior with
complex exponents near phase transitions; they also appear quite
generally in problems related to random-walks or classical diffusion
on self-similar systems or fractals, where the fractal dimension turns
out to be complex~\cite{hughes81,griffiths82,derrida84,sornette96,%
  akkermans09}. As such, they naturally occur in quasiperiodic systems
like the Fibonacci quasicrystal, where the spectrum of energies is
multifractal. This has been observed in quasiperiodic Ising
models~\cite{Andrade00}, and in specific heat studies of the
tight-binding Fibonacci model~\cite{vallejos98,daSilva01,oliviera05,%
  Mauriz01,Mauriz03}.  

In specific heat studies, where log-periodic oscillations appear in
temperature, the object of calculation is the partition function
\begin{equation}
  \label{eq:partition}
  Z(\beta_{\rm T}) = \sum_k e^{-\beta_{\rm T} E_k},
\end{equation}
where $\beta_{\rm T}$ is the inverse temperature; whereas in problems
of diffusion\cite{akkermans09}, the object of calculation is the
so-called heat kernel of the diffusion equation
\begin{equation}
  \label{eq:kernel}
  Z(t) = \sum_k e^{-E_k t}.
\end{equation}
In both cases, when expressed in integral form, these are \emph{Laplace
transforms of the density of states}.  To see the relation to the
present problem of quantum dynamics, we expand the expression for the
survival probability~(\ref{eq:survdef}) in eigenstates of the
Hamiltonian,
\begin{equation}\label{eq:survexp}
  S_n(t)= \left|\sum_k\bra{n}
    e^{iHt}\ket{\psi_k}\braket{\psi_k}{n}\right|^2 = \left|
    \sum_k\left|\psi_k(n)\right|^2 e^{i E_k t} \right|^2.
\end{equation}
Thus, what we are calculating is the magnitude-squared of the
\emph{Fourier transform of the density of states}, weighted by the
overlap of each eigenstate with the initial state $\ket{n}$. 

Equation~(\ref{eq:survexp}) is easily calculated in the periodic limit
$T\to1$ of our model, where the eigenstates are Bloch function, with
$\left|\psi_k(n)\right|^2=1/N$ independent of $n$, and the eigenvalues
form a single band with dispersion $E_k=\cos(k)$. This yields
\begin{equation}\label{eq:survperiodic}
  S_{T=1}(t) \propto \left|\int_{-\pi}^\pi  {\rm d}k e^{i t \cos k} \right|^2
  \propto J_0^2(t),
\end{equation}
where $J_0(t)$ is the zeroth order Bessel function of the first kind,
which is known to decay asymptotically as $t^{-1/2}$. This is
consistent with the expected ballistic dynamics of Bloch electrons in
a periodic crystal. The time average of Eq.~(\ref{eq:survperiodic}) is
approximately the behavior calculated numerically for $T=1$ on a
233-site approximant, and shown in Fig.~\ref{fig:MaxSTimeAv}. Note
that the fine wiggles that are barely seen at very early times are
associated with the first few zeros of the Bessel function, and are
not log-periodic oscillations.  Log-periodic oscillations appear for
$T>1$ as a result of the multifractal nature of the Fibonacci
spectrum, for similar reasons as in the calculation of the partition
function~(\ref{eq:partition}) \cite{Mauriz01,Mauriz03}. A more
detailed analysis of Eq.~(\ref{eq:survexp}) for $T>1$ is required to
quantitatively characterize the spectral properties of the
log-periodic oscillations, which we have discovered here by numerical
means.

Log-periodic oscillations in the specific heat of Fibonacci
quasicrystals near zero temperature are very difficult to observe
experimentally.  The discovery of log-periodic oscillations in the
quantum dynamics of wave-packets in Fibonacci quasicrystals should
open new possibilities for the actual experimental observation of this
unique phenomenon. A particular realization could be in optical
experiments that allow one to observe the dynamics of wave-packets
within 1-dimensional~\cite{lahini09}, as well as
2-dimensional~\cite{freedman06}, photonic quasicrystals. Thus, we hope
that our numerical observations here will stimulate further studies,
analytical and experimental alike.

\section*{Acknowledgments}

We wish to thank Amnon Aharony for fruitful discussions and for
suggesting that the ``wiggles'' in our numerical data might be real.
We also thank Eric Akkermans and Eli Eisenberg for helpful
discussions.  This research is supported by the Israel Science
Foundation through Grant No.~684/06.

\singlespacing
\small
\bibliographystyle{unsrt}
\bibliography{fibonacci}

\end{document}